\documentclass[review]{elsarticle}

\usepackage{lineno,hyperref}
\journal{Computer Networks, Elsevier}

\begin{document}

\begin{frontmatter}

\title{An Integrated Approach for Energy Efficient Handover and Key Distribution Protocol for Secure NC-enabled Small Cells}


\author[1,2]{Vipindev Adat Vasudevan \corref{correspondingauthor}}
\author[3]{Muhammad Tayyab}
\author[4]{George P. Koudouridis}
\author[4]{Xavier Gelabert}
\author[1,5]{Ilias Politis}

\address[1]{Wireless Communications Laboratory, University of Patras, Greece}
\address[2]{Universidade de Vigo, Vigo, Spain}
\address[3]{Department of Communications and Networking, School of Electrical Engineering, Aalto University, Espoo, Finland}
\address[4]{Huawei Technologies Sweden AB, Kista, Sweden}
\address[5]{Systems Security Laboratory, University of Piraeus, Greece}

\cortext[correspondingauthor]{Corresponding author: Vipindev Adat Vasudevan (vipindev@ece.upatras.gr)}

\begin{abstract}
Future wireless networks must serve dense mobile networks with high data rates, keeping energy requirements to a possible minimum. The small cell-based network architecture and device-to-device (D2D) communication are already being considered part of 5G networks and beyond. In such environments, network coding (NC) can be employed to achieve both higher throughput and energy efficiency. However, NC-enabled systems need to address security challenges specific to NC, such as pollution attacks. All integrity schemes against pollution attacks generally require proper key distribution and management to ensure security in a mobile environment. Additionally, the mobility requirements in small cell environments are more challenging and demanding in terms of signaling overhead. This paper proposes a blockchain-assisted key distribution protocol tailored for MAC-based integrity schemes, which combined with an uplink reference signal (UL RS) handover mechanism, enables energy efficient secure NC. The performance analysis of the protocol during handover scenarios indicates its suitability for ensuring high level of security against pollution attacks in dense small cell environments with multiple adversaries being present. Furthermore, the proposed scheme achieves lower bandwidth and signaling overhead during handover compared to legacy schemes and the signaling cost reduces significantly as the communication progresses, thus enhancing the network’s cumulative energy efficiency.
\end{abstract}


\end{frontmatter}

%\linenumbers

\section{Introduction} \label{intro}
%Para 1: 5G, Small Cells (SC), D2D communication, Mobility
The exponential growth in the number of digital devices and the data traffic generated by them has forced a major upgrade in communication technologies. The cellular communication technology is changing in its entirety and the fifth-generation (5G) of wireless technology being rolled out in experimental studies shows the paradigm shift towards future networks. The 5G and beyond (B5G) networks are expected to have significant changes in the underlying technologies as well as in the system architecture \cite{5Gsurvey,gupta2015survey}. Focusing on providing a high data rate for a dense network of mobile users, small cell ultra-densification, by placing base stations (BSs) in close proximity, becomes one of the major structural changes in future cellular networks. This reduces the power requirements of BSs without compromising the quality of services by reducing the area and the number of users to be served by a single BS. On the other hand, device-to-device (D2D) communication \cite{mumtaz2014direct} will supplement the paradigm shift from a BS-centric approach to a more user-centric approach into practice. The operation of B5G networks will also be supported by different novel technologies such as, e.g., blockchain, artificial intelligence, network coding, and software-defined networks which may not be the subject of standardization but still play pivotal roles for the network operation automation and the supported applications in the 5G era. In particular, in a cooperative environment of small cells and D2D-enabled devices, network coding (NC) provides exciting prospects to achieve high network throughput and efficient bandwidth usage. Furthermore, NC's underlying principle, allowing the intermediate nodes in a hop-by-hop network to code the packets on the fly, will also enhance the reliability in a wireless environment with low energy requirements \cite{SECRET2017secret}. Despite NC-enabled small cells can match multiple use cases in the B5G environment \cite{vieira2017network}, it also suffers from new security challenges. Specifically in this work, we draw our attention to the pollution attacks in an NC environment in combination with the increased handover (HO) signaling rates due to small cell deployments. These are briefly described in the following paragraphs.  

%Para 2: Network Coding (NC) enabled SC \cite{SECRET2017secret}, Security and Key management 
%The NC enabled mobile small cells can match multiple use cases in the B5G environment \cite{vieira2017network}, but also suffers from different security challenges. 
In essence, NC enables the intermediate nodes\footnote{The term \emph{node} in the context of NC has a wide meaning and could describe any communication node in a communication system, including source, destination, and intermediate nodes. When necessary, a further distinction will be made in the article.} in the network to code the packets they receive and send the coded packets (these packets in transition are referred as in-transit packets) to the outgoing links. This unique property of NC also introduces a security vulnerability in the system. An adversary can exploit this ability to inject a corrupted packet into the packet flow. This kind of attack, where an adversary node injects a corrupted packet, therefore \emph{polluting} the communication in a network, is called a pollution attack in NC scenarios. If a polluted packet is introduced to the network and not detected, all subsequently coded packets containing the polluted packet will also be corrupted. Unsurprisingly, this contagious property of pollution attacks makes it one of the major challenges in NC implementations. Since the in-transit packets are coded by participating nodes, generic integrity schemes, such as hashes, signatures, or message authentication codes, will not be sufficient to prevent pollution attacks. To this end, homomorphic Message Authentication Code (MAC)-based integrity schemes are used to enable secure NC \cite{agrawal2009homomorphic}. However, the MAC-based integrity schemes require some symmetric keys to be shared between the source node and receiving nodes. Most of the integrity schemes suggest having a pre-distribution of these keys and based on specific key distribution models such as c-cover free system \cite{canetti1999multicast}. These key distribution schemes play a crucial role in protecting the integrity scheme from colluding adversaries. Essentially, traditional key distribution schemes are used to distribute a set of keys in such a way that no node will have all the set of keys the source node has used to create the tags. However, achieving such key distributions with constraints depending on the number of participants in the network can be troublesome in a dense network of mobile nodes. Furthermore, in the case of a mobile network, the neighborhood of nodes can be volatile making it even more difficult to achieve practically. Thus moving towards the future networks, the idea of key distribution for secure NC-enabled environments requires significant advancements. 

There are a few fundamental requirements on the key distribution schemes for secure NC applications compared to traditional key management and distribution protocols. These keys are essentially used to ensure the integrity of the in-transit packets rather than authenticating the nodes in the network. Further, they should also be able to ensure integrity check over RLNC coded packets, thus requiring homomorphic properties for the integrity scheme. Even though the MAC-based integrity schemes require symmetric keys to be shared in the network, these are not pairwise keys as in a private key cryptosystem. The same set of keys needs to be shared with multiple participants, making it a special case of symmetric key distribution for NC applications. In such an arrangement, the secret key is not unique for a particular node, but multiple participants may use the keys from the same set for creating their tags while acting as the source node. Thus the keys can not be used to differentiate the participants in the proposed environments. Further, the key distribution does not happen pairwise, so when a new node is entering the network, there is no need to be given a new unique key, but be provided access to the existing key set in the network. This is also a special case of the proposed key distribution for secure NC-enabled environments which also has implications in the handover mechanism which involves the movement of nodes.

The major shortcomings of the existing key distribution schemes in secure NC applications are the lack of scalability and suitability to a mobile environment. The two major variants of the key distribution for the MAC-based integrity scheme were present in \cite{zhang2011padding} and \cite{esfahani2017efficient}; both based on approaches mentioned in \cite{canetti1999multicast}. However, both these approaches have a limit on the number of adversaries that can be successfully prevented which will depend on the number of keys in the network. This will also impact the scalability of the network since the probability of colluding attackers will increase with the increase in the number of nodes and a larger number of keys will be required to accommodate more nodes. Furthermore, a pre-distributed key management scheme is not ideal for a mobile network. In addition to the aforementioned security challenges, small cells introduce additional mobility challenges in terms of an increased number of HOs and, consequently, HO signaling, owing to decreased cell sizes. In this respect, with regards to HO signaling, proper key sharing during HOs is fundamental for the participating nodes to ensure security and harness the benefits of NC properties. As the participants may move from one small cell to another, it may require new keys. Updating an incoming participant with the set of keys used in the target cell may be performed as part of the handover procedure. To facilitate this, the existing schemes may have very high overheads to share the required keys. As a result, the mobility requirements become more demanding both in terms of reducing HO failures and HO-related signaling, altogether calling for a re-design of current HO procedures.

% contribution paragraph
In this paper, we jointly address the aforementioned challenges by proposing a blockchain-enhanced key distribution scheme to ensure the integrity of packets in NC-enabled small cell environments. The aim is to address the problem of key sharing and propose a solution using a blockchain-like distributed ledger that satisfies the requirements for low latency and minimum processing capacity \cite{adat2019towards}. This scheme is further aided by an uplink reference signal (UL RS) based HO scheme \cite{tayyab2020ULHO} to address the problem of high HO signaling overhead in small cell deployments which encompasses a higher number of key exchanges. We further analyze the key management overheads during HOs to ensure secure NC in the small cell environment. We combine the UL RS based HO (UL-HO) scheme and the blockchain-based key sharing approach and propose a blockchain-based security and HO controller (BSH controller) to ensure a smooth and secure HO process for NC-enabled mobile small cells. The blockchain-based key management reduces the number of signal exchanges during HOs compared to traditional approaches, and the UL RS based HO schemes further assist towards reducing the latency incurred by the blockchain verification. This proposed approach is studied and compared with other traditional schemes in terms of signaling overheads and different security parameters. The extensive analysis and discussions on the proposed approach show the following novel contributions:
\begin{itemize}
    \item A distributed key management scheme suitable for highly mobile secure network coding enabled environment.
    \item The proposed architecture ensures high security against pollution attacks in dense small cell environments with low bandwidth overhead even if multiple adversaries are present.
    \item Energy efficient handover with low signaling overhead compared to legacy schemes.
    \item Signalling cost reduces significantly as the communication progresses, thus enhancing the network’s cumulative energy efficiency.
\end{itemize}

%Para 4: Paper structure 
The remaining sections in the paper are arranged as follows. A literature review of different MAC-based integrity schemes for secure NC with a focus on key management protocols and a review of legacy HO schemes are presented in section \ref{literature}. Section \ref{System Model} describes the system model of the NC-enabled small cells and how the BSH controller helps to secure HOs. The simulation scenarios and analysis of the proposed approach in terms of bandwidth and signaling overhead, as well as different security parameters, are presented in section \ref{analysis} followed by section \ref{conclusions}, which concludes the paper.

\section{Related Work} \label{literature}
This section discusses preliminary details and state of the art related to secure network coding and mobility challenges in the 5G and small cell environment.  

\subsection{Network Coding Preliminaries}
The idea of NC was introduced as a max-flow min-cut theorem for network information flow in a multicast scenario by Ahlsewede et al. \cite{ahlswede2000network}, noting that coding at intermediate nodes offered substantial benefits compared to simple routing or forwarding in terms of bandwidth efficiency. Subsequently, a similar study extending the idea of NC to directed acyclic graphs to find the optimal max-flow min-cut bound was presented in \cite{li2003linear}. It was proposed therein that the optimum can be achieved when linear coding is applied to the multicast scenario using coefficients from a finite field. Thereafter, it was proposed that randomly chosen coefficients could be used to perform linear NC and still decode packets efficiently, using matrix theory, when a sufficient number of linearly independent coded packets was received \cite{ho2003randomized,ho2006random}. As a result, random linear network coding (RLNC) paved the way for using NC in dynamic and volatile environments such as wireless networks, with \cite{torres2015network, SECRET2017secret} identifying some prospects of using RLNC in B5G. The interested reader is referred to a detailed study of different dimensions of NC theory and principles presented in \cite{bassoli2013network}.

\subsection{Secure Network Coding-Enabled Small Cells} \label{SNC Review}

Pollution attacks are considered as one of the most dreadful attacks in NC-enabled environments. They are difficult to mitigate and can spread across the network reducing the throughput significantly. Generic integrity schemes fail to ensure integrity in NC-enabled environments since the packets are recoded during transmission \cite{adat2020survey}. However, homomorphic MACs can be successfully used to ensure the integrity of coded packets. Such MACs that provide homomorphic properties over RLNC operations are first proposed in \cite{agrawal2009homomorphic}. In MAC-based integrity schemes, message authentication codes are created over the packet by the source node and attached as tags to the packet. These tags can be verified by the receiving nodes if they possess the keys used to create the tags. If a malicious user gets the key, it can pollute the packet and create a valid tag so that the polluted packet will not be detected and thus making the integrity scheme void. This issue has been addressed in different integrity schemes by different key distribution models. Another challenge in the MAC-based integrity schemes is the tag pollution attack. In tag pollution attacks, the malicious user will pollute tags of a genuine packet which will result in the discarding of that packet by a benign receiver. This will also reduce the throughput of the network. In this subsection, we review three prominent MAC-based integrity schemes that tackle both data pollution and tag pollution attacks and provide an overview of the key distribution schemes they employ. 

MacSig \cite{zhang2011padding} was one of the initial integrity schemes which addressed the issue of both data pollution and tag pollution attack. MacSig uses homomorphic MACs to ensure data integrity and a homomorphic signature to protect from the tag pollution attacks. A double random key distribution scheme is used in MacSig to ensure safe key distribution. In practice, more than one tag will be used to achieve enhanced security using a set of keys. An attacker will have to acquire all these keys to bypass the integrity check. In double random key distribution, every participating node is initially provided with a set of random keys as in \cite{canetti1999multicast}. Then, while initiating the transmission, the source chooses $l$ random keys from the key set it holds and uses these keys for tag creation. Nodes that have at least one of these $l$ keys can verify the integrity of the packets during the transmission. If a receiving node does not have any of the keys used to create the tags, it will not be able to verify the integrity of the packets. However, even if the receiving node holds most keys, there is a slight chance of a polluted packet to escape the integrity check. This implies that the probability of a polluted packet passing the integrity check at a receiving node with $l'$ matching keys depends on both the number of matching keys and the field size of the NC operations ($q$), and is equal to $1/q^{l'}$.

Another integrity scheme, which addresses both data pollution and tag pollution attacks using a different key distribution and referred to as homomorphic MAC (HMAC) scheme, is proposed in \cite{esfahani2017efficient}. Therein, Esfahani et al. employed a key distribution model based on c-cover free set systems \cite{canetti1999multicast}. In this proposed scheme they use MACs to ensure data integrity and a so-called D-MAC with a signature to protect against tag pollution attack. The keys are distributed in a random fashion as per the principles of c-cover free set systems. In HMAC, two sets of keys are shared with the source node and different subsets of them are shared with the remaining nodes. To ensure the key distribution is $c$ collision-resistant, the number of keys at the source node should be at least equal to $c$ times the number of keys each participating node has. More precisely, in HMAC, every participating node, other than the source node, will have only one key from the key set of the source. This limits the integrity check at each node to a single tag verification which allows a polluted packet to pass through the integrity check with a probability of $1/q$. There are also recent schemes in the literature such as \cite{parsamehr2019novel,parsamehr2020idlp} which use the same cover free set systems for key distribution. However, these schemes also require an additional swapping vector to be secretly shared between the participating nodes for the purpose of enabling null-space based integrity schemes proposed in these works. This is an added overhead over the c-cover free key distribution. Another type of integrity scheme was presented by Lawrence et al. in \cite{lawrence2021hmac,lawrence2021computationally}. However, this scheme is also having key distribution based on cover free systems, with an additional arrangement to reduce the bandwidth by carefully choosing the first two keys. This is additional to the conditions of cover free systems. This also means that the recent schemes have more or equivalent complexity to the c-cover free system described in \cite{zhang2011padding}. Thus, even though there are recent schemes using the key distribution scheme, we compare our approach with the basic and most efficient key distribution process following the c-cover free system presented in \cite{zhang2011padding}. 

Finally, a blockchain-based integrity scheme is proposed for next-generation small cells in \cite{adat2018blockchain}. This integrity scheme uses homomorphic MACs to ensure data integrity and shares these MACs with other participating nodes through a blockchain overlay of small cells. This integrity scheme allows all the participating nodes to have the complete set of keys used by the source nodes and verification of all the tags attached to the packets. This ensures that if $L$ keys are used to create the tags, every receiving node can verify all the $L$ tags, limiting the probability of a polluted packet passing through a benign node to $1/q^L$. The blockchain can also be used to share the keys required for MAC verification. To this end, \cite{adat2019towards} studies how the blockchain-based key sharing scheme helps to extend the integrity scheme to a network of multiple NC-enabled small cells and support inter-cell mobility for users.

\subsection{ Key Management Schemes for NC-Enabled Small Cells}
\label{key management review}

As discussed in the introduction, secure NC-enabled small cells use HMAC based integrity schemes to prevent pollution attacks which require symmetric keys to be distributed among the participating nodes. Increasing the complexity of the required key management scheme, a source node will have to share a secret key with different nodes in the network to ensure the integrity checks are possible with intermediate and destination nodes. In other words, even though the keys are symmetric, they are not pairwise, rather the keys used by a source node should be available to the nodes that are involved in the information flow started by the source node. This requirement of HMAC-based integrity schemes is different from the general authentication schemes and the key management scheme also has some additional requirements as described in the introduction. The proposed approach should be scalable and support a highly dense mobile environment spread across a large geographical area and be suitable for the small cell scenario in 5G and beyond networks \cite{lu2008framework,khan2011energy,tian2020blockchain,de2021distant}. Towards this end, we are looking at different decentralized integrity schemes to distribute the secret keys used for HMAC-based integrity schemes to the participating nodes.

The key distribution schemes can be broadly classified into centralized, partially distributed, and fully distributed approaches \cite{mark2019key}. The centralized approach largely depends on the availability of a central trusted third party (TTP) for generating the keys, distributing them to the participants, and managing the key life cycle. However, the availability of such a TTP over a large network throughout its operation has been a point of concern from the time of mobile ad hoc networks. This has led to the idea of decentralized schemes to distribute the keys. A partially distributed scheme distributes the duty of key distribution to a proper subset of network nodes and these nodes will manage the key life cycle for the remaining participants. For example, in a small cell environment, the small cells may be provided with the additional responsibility of key management, ensuring these nodes can be trusted in the system. Fully distributed schemes also exist in the literature where the trust is evenly distributed across all the participating nodes. This work focuses on a partially distributed approach of symmetric key sharing where the small cells are entrusted with the key distribution.

de Ree et al. \cite{mark2019key} discuss relevant key management schemes on the backdrop of 5G ad beyond networks and discuss the requirements and favorable characteristics of key management schemes for a network of mobile small cells. The symmetric key management schemes can be broadly divided into key pre-distribution, key distribution, and key agreement. The key agreement schemes require a set of participating nodes to interact among themselves to establish a shared symmetric key. This type of schemes does not require any centralized trusted third party to manage the keys but they are also not very stable against network topological changes. Furthermore, it will take more time and message exchanges to reach an agreement on the shared secret which can increase the communication and latency overheads in key management. Such schemes also rely on specific secure routing infrastructure to avoid man-in-the-middle attacks while key agreement. 

Both key pre-distribution and distribution classes are organized by a trusted third party called the Key Distribution Center (KDC). In the first set of schemes such as \cite{blom1984optimal,matsumoto1987key,du2005pairwise,ramkumar2005efficient}, the pairwise keys are distributed by the KDC in the network initialization phase and the KDC will be offline afterward. This restricts new nodes entering the network after the initialization phase from getting proper keys to communicate with existing nodes in the network. Most of the previous integrity schemes \cite{zhang2011padding,esfahani2017efficient} addressing pollution attacks used key pre-distribution approaches \cite{canetti1999multicast} since they focused on static networks. However, the pre-distribution models are resilient against dynamic network environments such as mobile networks making them not the ideal choice for future wireless networks.

The class of key distribution schemes differs from the pre-distribution schemes mainly on the availability of the KDC during the network lifetime. In key distribution schemes, the KDC will be online during network deployment and any node can interact with the KDC to obtain a temporary key to communicate with another node in the network. Thus each node will only store a long-lasting symmetric key to interact with the KDC and obtain the temporary key on-demand for interacting with other nodes in the network. Several key distribution approaches designed for mobile ad-hoc networks \cite{wong2000secure,liu2012key,alam2014secure} employ a centralized KDC which creates the risk of a single point of failure as well as restricts the capacity to support a dense mobile network. In a dense mobile network, the availability of the centralized KDC may be limited due to different factors such as communication range, congestion, unknown network topology, increased mobility etc.  

Considering the above aspects, this paper proposes a distributed key management protocol for secure NC-enabled environments. The idea of using decentralized key management schemes to suit the requirements of mobile networks in the 5G and beyond era has got significant research attractions in the recent years \cite{lu2008framework,khan2011energy,tian2020blockchain,de2021distant}. The proposed key distribution protocol employs the distributed ledger hosted by the BSH controllers to efficiently distribute the keys in the dynamic environment. A detailed description of the scheme is presented in the following section \ref{System Model}. It is to be noted that the specific key management and distribution protocols that are discussed in the scope of this paper are used for efficient integrity schemes to prevent pollution attacks and not necessarily be used for authentication of the participating devices. The key management for authentication of devices in such environments may be performed with any of the previously discussed approaches in the literature. 

\subsection{UL Handover Scheme For Small Cells} \label{HO Review}
%one small paragraph related to UL HO scheme (literature review)
As discussed in previous subsections, the security of the NC-enabled small cells depends on the integrity schemes and the key distribution. In a mobile environment, the key distribution should also be a part of the HO protocol and the users requesting a HO should also be provided with the required keys to ensure the integrity of packets it receives.
In this setup, the adoption of the UL-HO scheme for small cell environments brings substantial improvements with respect to the legacy DL RS based HO scheme as noted, e.g., in \cite{tayyab2020ULHO,xavier_gelabert_uplink_2016}. Therein, UL RSs are transmitted by the User Equipments\footnote{A mobile device in 3GPP terminology. Both terms will be used interchangeably.} (UEs) which are then received at the serving as well as the nearby BSs for posterior measurement and processing. A central network controller collects and further processes these measurements and triggers a HO if deemed necessary, eliminating the subsequent measurement reporting phase of the legacy HO schemes. Noteworthy, as noted in  \cite{tayyab2020ULHO}, the UL-HO scheme reduces the HO rate due to a reduction in HO failures and unnecessary HOs (i.e. ping-pongs). As a result, the number of security key exchanges during HO also decreases. 

In addition, research undergone in \cite{tayyab2020ULHO,xu2019uplink,Qualcomm3GPP,Sony3GPP} further shows that the UL-HO scheme is more suitable for future small cell deployments in terms of reducing UE power consumption. Another benefit from implementing  UL RSs in small cell networks is that enhanced location and UE tracking can be effectively and accurately achieved \cite{hakkarainen_high_efficiency_2015}. This opens the possibility of new prospects in location-aware services and network optimization, for example allowing the prediction of the UEs position and enabling proactive Radio Resource Management (RRM) strategies. This facilitates, for example, that any necessary information can be pre-fetched in advance, thus contributing to optimized end-to-end latency \cite{hakkarainen_high_efficiency_2015}. A direct application of this may be to proactively initiate the key exchange once a handover is deemed predictable.  

\section{System model} \label{System Model}

This paper considers a small cell environment where the participating mobile devices (i.e. UEs) are using NC for D2D communications. More specifically, mobile devices use RLNC for D2D communication and these mobile devices are connected to the core network through small cell hotspots or BSs. Further, we consider the BSs in the network have an inbuilt functionality termed as blockchain-based security and HO (BSH) controller. These BSs can potentially be any equipment (e.g. a pico/micro/macro BS, mobile devices, etc.) that have both the computational capacity and the  capability  of providing cellular services over a small area, i.e. a small cell. We term such a cell as a BSH-enabled cell, or BSH-cell. These BSH controllers form a blockchain overlay that serves as a secure distributed ledger to support the integrity scheme and the key management protocol presented in this paper. All mobile devices, referred also as nodes in the NC and blockchain context, are connected to at least one BSH-controller through a secure control channel, as shown in fig. \ref{fig1}. Particularly, this work focuses on the key exchanges that happen during the HO process when a node that is part of the NC-enabled communication network in one BSH-cell moves to another BSH-cell in order to be part of the NC-enabled communication in the destination BSH-cell. 

\subsection{Blockchain-based security and handover controller} \label{blockchain}
%We describe the network model in which small cell heads will act as blockchain nodes and all users capable of NC enabled D2D communication.

\begin{figure*}
\includegraphics[width=12cm]{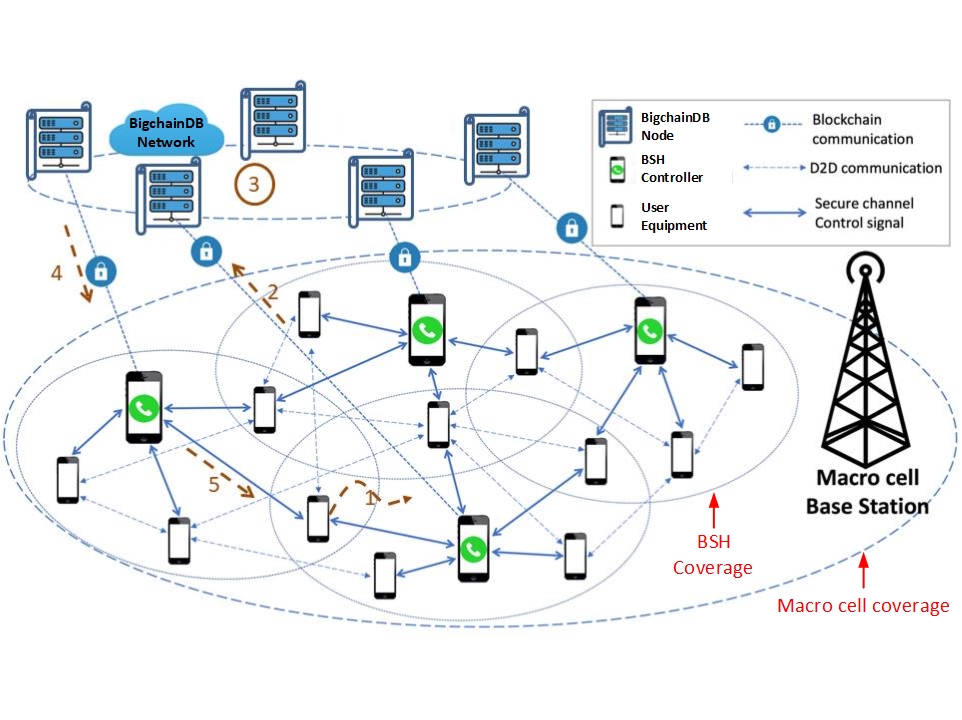}
\centering 
\caption{Key exchange through the blockchain overlay of BSH controllers. Numbered arrows represent different steps in the HO procedure.}
\label{fig1}
\centering
\end{figure*}

Blockchain is a distributed and decentralized ledger where data is stored as immutable and secure blocks \cite{underwood2016blockchain}. In the proposed scheme, we consider a blockchain implementation enabled by bigchainDB in which the BSH controllers are considered as the bigchainDB nodes, as shown in Fig.~\ref{fig1}. The bigchainDB is a distributed database with blockchain features. It ensures immutable storage of information as connected blocks and also enables query based data retrieval from this ledger. Each BSH-cell comprises a security domain that has its own keys to ensure the integrity of the RLNC communication. Any UE that exists in a given BSH-cell are considered belonging to the same security domain. In addition, all HO and security management of all UEs within a BSH-cell are taken care of by the BSH controller as an internal HO. All HO and security management of users moving across BSH-cells are taken care of by the BSH controllers of the corresponding BSH-cells in a distributed manner using the blockchain. To this end, the HO is also performed in a distributed manner between BSH controllers of different security domains and their corresponding BSH-cells. Even though UEs are not part of the bigchainDB network, they can send information to be stored in the bigchainDB ledger to the nearest BSH controller and it will be included in the next candidate block, i.e. a block which will eventually be stored in the bigchainDB. Similarly, UEs can query the nearest BSH controller for the information stored whenever necessary. 

For the integrity scheme, once the tags are created at the source node, these tags are also uploaded to the ledger. Any receiving node can verify the packet integrity by comparing the tags received via the communication channel and the tags fetched from the BSH controller. This ensures that the proposed scheme is secure against both data pollution as well as tag pollution attacks \cite{adat2019blockchain}. However, this integrity scheme also requires a set of keys. The blockchain overlay can also be used to ensure that these keys are properly shared with the users that are part of the network even if they move from one BSH-cell to another one.

\subsection{Proposed Blockchain based Key Management Scheme} \label{key management}
%The distributed key management scheme will be explained here. Initially (in the pre-communication phase), the keys will be distributed in SC level (or cluster level). For a user moving from one SC to another, it will need the new set of keys to participate in the communication there. We enable this key sharing using the blockchain overlay explained in section \ref{blockchain}. 

In the MAC-based integrity scheme, the source node creates some security tags over the native packets. These tags are attached to the packets before transmission as well as shared within the network through the blockchain overlay implemented using bigchainDB \cite{bigchainDB2}. A receiving node can verify the authenticity of the tags received through the communication channel by comparing it with the corresponding tags stored in the bighchainDB and then verify the integrity of the packets by recreating the tags using the shared secret key over the received packets. The integrity scheme in \cite{adat2018blockchain} considers a very strong adversary condition where the adversary may have access to all the keys used by the source node but still not be able to bypass the integrity check. This simplifies the key distribution process since all the participating nodes can have access to all the keys used to create the tags. 

However, in a dense small cell environment, distributing keys to every participating node directly by a single key distribution entity may not be practical. Further, different small cells may belong to different BSH controllers and hence different security domains depending on their location, service provider, and application background. Thus the keys for the integrity scheme may be shared within a BSH-cell. Every node (e.g. a UE) in a particular BSH-cell shall have access to the set of keys used in that cell so that it can verify the integrity of packets being generated and transmitted in that BSH-cell. However, any new node entering a new BSH-cell should also acquire this set of keys as part of their authentication to be part of the NC communications in the BSH-cell. This is also applicable to a node moving from one BSH-cell to another one. This is enabled by the distributed overlay of the BSH controllers similar to the concept in \cite{adat2019towards}. 

The key management happens in two major phases. The initial phase, which can be termed as the pre-distribution phase, happens during the initialization of the network. Each BSH-controller will have its own set of keys and these keys are shared with the nodes authenticated under the particular BSH controller. The second phase of the key management happens dynamically when a new node is entering the network or a node is moving from one BSH-cell to another one. At this stage, the node will have to be authenticated by the BSH controller and given access to the set of keys used for integrity check in that BSH-cell. If the node is moving from one BSH-cell to another, the set of keys used in the target BSH-cell should be shared with the incoming node during the HO process. In our proposed approach this is achieved through a distributed ledger, contrary to traditional approaches where the host and target controllers communicate directly during the HOs. Every BSH controller is part of the blockchain overlay as described in \cite{adat2019blockchain}. This ensures that if a node has entered a target BSH-cell, the keys associated with the particular cell will be stored in the distributed ledger and the next incoming node will not have to repeat the same procedure. The serving BSH controller of this new incoming node has access to the distributed ledger and can directly share the set of keys used in the target BSH-cell with the mobile node. This reduces the number of signal exchanges required for the key distribution during HO. In this paper, we focus on the impact of this key sharing scheme on the HO procedure.

Whenever a node request for a HO, the target BSH controller will upload the set of keys that are used in that cell to the blockchain as a candidate block. As per the requirements of HMAC-based integrity scheme proposed in \cite{adat2018blockchain}, a key is a vector defined in the finite field $F^{n+1}_q$ where $F^{n+1}_q$ is the field of RLNC operations, $n$ is the native packet size, and $q$ is the field size. The blockchain will validate the candidate blocks during the collection period between two block verifications and once the candidate block is validated, the source BSH controller will also have a copy of the verified block through the consensus achieved in the blockchain and it can share the required keys with the node that is undergoing HO. This validation of blocks is the major activity during the dynamic key management phase. This key sharing is performed along with the authentication process during HO since not all nodes in the network needs to hold the keys for all the small cells. Only those mobile nodes which moves to a new small cell may get it from the BSH controller and it is done as part of the HO signalling. A step by step description of the HO process is shown in fig. \ref{fig1} and discussed in subsection \ref{Uplink HO}. Noteworthy, all the nodes requesting HOs between the validation of two consecutive blocks will be served with the keys once the new block is updated to the blockchain.

Once a BSH controller stores its key set to the blockchain, any further incoming node (via HO) will have easy access to that specific set of keys. Since the blockchain is an expanding list of records, a verified block will never be modified and always be available to the participating nodes. Further, it is a distributed and decentralized ledger so that, once a block is verified and added to the chain, it will be available not only to the BSH controller which initiated the request but with all the BSH controllers which are part of the blockchain. Thus any further node requesting a HO can be served with the keys directly from the source BSH controller. Further, all the candidate blocks during a specific collection period will be updated to the blockchain in a single block verification. This reduces the overhead of signaling considerably in a dense network of mobile nodes. A detailed analysis of the advantages of this proposed scheme is presented in \ref{analysis} and the HO scheme is explained in \ref{Uplink HO}.

\subsection{Uplink Reference Signal based Handover Scheme for Small Cells} \label{Uplink HO}
\begin{figure}
\centering
\includegraphics[width= \columnwidth]{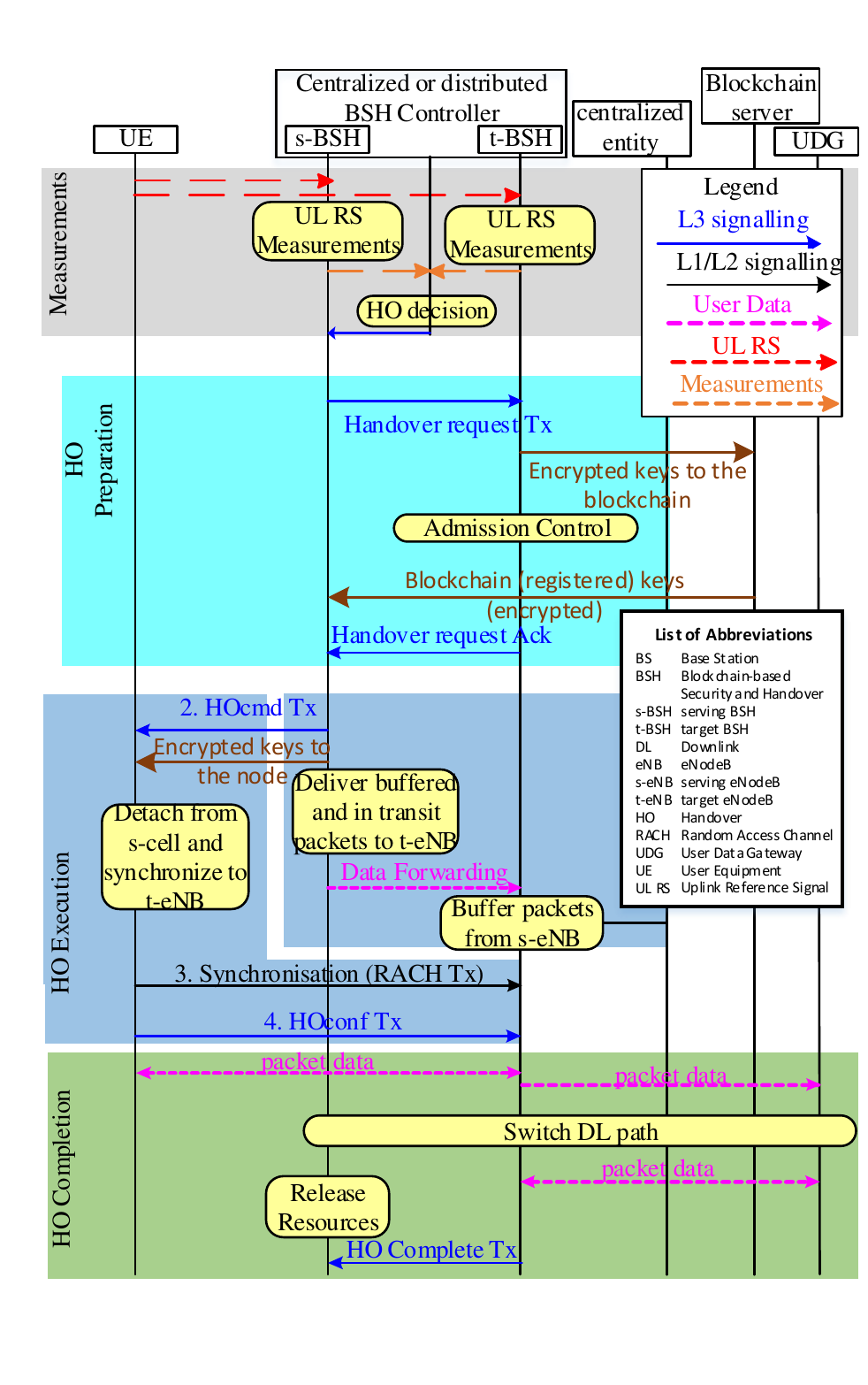}
\caption{Uplink Reference Signal based HO (UL-HO) scheme with blockchain-based key management.}
\label{fig2}
\centering
\end{figure}

An overview of the UL-HO scheme with a blockchain-based key management protocol is presented in this subsection, according to Fig. \ref{fig2}. The actions related to the blockchain-based key management mentioned in this subsection are the actual procedures followed during the dynamic key management phase, explained as different phases during the HO process. In the measurement phase, the UE transmits UL RSs reaching the serving BS (s-BS) and  nearby BSs, a subset of which are candidate HO target BSs (t-BSs). Both the s-BSH controller and the t-BSH controllers perform the UL RS measurements and either communicate this information to each other, assuming that each of the involved BSs implements a BSH cell and employ a BSH controller, or to a central network BSH controller, where the s-BSs and t-BSs belong to. In the latter case, the centralized network BSH controller processes these measurements to determine which BS shall serve the UE, as shown in Fig.~\ref{fig2}. For this purpose, the network BSH controller checks if the UL RS received power (UL RSRP) of any of the candidate t-BSs is higher than the s-BS by an offset called UL-offset and this condition is maintained within a time defined by UL time to trigger (UL-TTT). When a suitable t-BSs is identified, the network BSH controller takes the HO decision and sends the potential t-BS information to the s-BS. In case s-BS and t-BSs employ their own BSH controllers, this decision is taken in a distributed manner among the corresponding s-BSH and t-BSHs controllers. 

In  the next phase, known as HO preparation phase, the s-BS sends the HO request to the t-BS. Upon successful admission control, the t-BSH controller shares the keys, where each key vector is defined in the finite field $F^{n+1}_q$, with the s-BSH controller either directly, in case of a distributed setup, or via the centralized entity where the network BSH controller belongs to. In LTE such a centralized entity may be the Mobility  Management  Entity (MME), while in 5G New Radio (NR) it may be the Access and Mobility Management Function (AMF). Once the key sharing is completed, the t-BS acknowledges the HO request to s-BS. 

Now the HO execution phase starts where the s-BS sends a HO command to UE. The UE starts accessing the t-BS through a random access procedure. Upon successful synchronization with the t-BS, the UE sends a HO confirm message to the t-BS. The s-BS shares the keys, where each key vector is defined in the finite field $F^{n+1}_q$ where $n$ is the packet size and $q$ is the field size, with the UE and releases the resources. In our approach, the field size is set to $2^8$ and packet size is 1024 bytes, unless specified otherwise. Finally, the t-BS sends a HO complete message to the s-BS to inform the success of the HO procedure when the DL data path is switched from the user data gateway (UDG) toward the t-BS. The UDG is defined as the serving gateway (SGW) in LTE and the User Plane Function (UPF) in NR. It is to be noted that all the communications related to the keys and other information are encrypted as per the standard security protocols. 

It has to be noted that in the case where the centralized entity and the network BSH controller are assumed, the key sharing step between the t-BS and s-BS may not be necessary. Only when the security domains are different for different BSH-cells, the HOs between different BSH-cells require key sharing for the integrity scheme. In the context of this paper, only such HOs are of interest. In addition, the key sharing as we described in \ref{key management} happens differently for the initial and subsequent HOs. For the initial HO to a particular t-BS, when the UE requests for a HO (Step 1 in Fig.~\ref{fig1}), the t-BS shares its key set to the blockchain server as a candidate block (Step 2). Once this block is verified and added to the blockchain (Step 3), every BSH controller including the s-BSH controller will receive this block (Step 4). This is performed as part of the admission control in the HO preparation phase, as shown in Fig.~\ref{fig2}. However, with the UL-HO scheme, possible HOs can also be predicted earlier to avoid delay in verifying the block and updating the blockchain. Then at the HO completion phase, this set of keys are shared by the s-BS to the UE requested the HO (Step 5). From the second HO request to the same t-BS, the blockchain already has the keys stored in it. This eliminates the requirement of any signaling in the HO preparation phase for further handovers and the key set can be sent to the UE at the HO completion phase by the s-BS, as shown in Fig.~\ref{fig2} and step 5 of Fig.~\ref{fig1}.

\section{Analysis} \label{analysis}

This section presents a performance evaluation of the proposed blockchain enhanced key management protocol for MAC-based integrity scheme combined with an UL-HO scheme. The performance of the proposed scheme is evaluated by means of simulations and compared to the other two prominent key management schemes, namely HMAC and MacSig. The evaluation analysis is performed in terms of security, bandwidth and signaling overhead. 

\subsection{Simulation Scenario and Assumptions} \label{simulation assumptions}

\subsubsection{Network Topology and Mobility Assumptions} \label{network assumptions}
\begin{figure}
\includegraphics [width = 0.9\columnwidth]{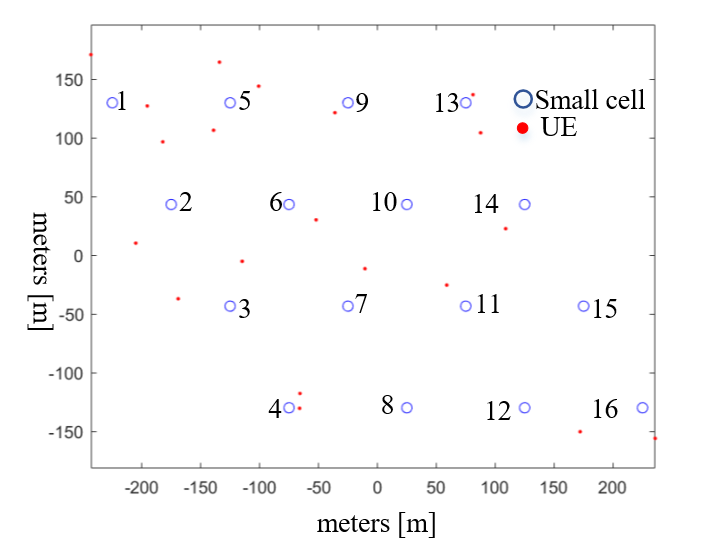}
\centering
\caption{Considered simulation scenario.}
\centering
\label{fig3}
\end{figure}

The simulation scenario is shown in Fig.~\ref{fig3} which contains sixteen BSH controllers with inter-site distance (ISD) of 100 m in a MATLAB based system-level simulator. The cell wrap-around feature is included to have fair interference conditions across the simulated area where a set of 20 UEs are randomly placed according to a uniform distribution. The UEs move at a fixed speed of 60 km/h with random directions [0°, 360°] to resemble the scenario with the UEs moving on private vehicles (i.e. cars). The UL-HO procedure is used with a UL RS periodicity of 160 ms, UL-offset of 1 dB, and UL-TTT of 32 ms. The UL RS periodicity refers to the rate at which a UE transmits its reference signals in the UL. UL-offset defines the difference in UL RS power conditioning the HO from a s-BSH to t-BSH, while UL-TTT defines the time-to-trigger the HO procedure. The reason for choosing low UL-offset and UL-TTT is because of keeping HO rate to its minimal level at high speed (see \cite{tayyab2019smallcell} for details of HO parameters selection for the DL-HO case).  Similarly, high UL RS periodicity is selected to reduce the rate of unnecessary HOs. Simulation assumptions and parameters are summarized in Table.~\ref{table1}. Further details regarding the simulator modeling, UL RS model, and the parameters used are covered in \cite{tayyab2020ULHO}.

\begin{table}
\centering
\caption{Simulation assumptions and main parameters.}
\label{table1}
\begin{tabular}{lll}
\hline
\hline
%\multicolumn{2}{c}{Item} \\
%\cline{1-2}
\textbf{Model}   & \textbf{Parameter} & \textbf{Value/Derivation} \\
\hline
Network   & Cells               & 16      \\
          & ISD                 & 100 m       \\
          & Interference        & cell wrap-around    \\
          & UEs                 & 20       \\
          & UE speed            & 60 km/h       \\
          & UE direction        & random from [0°, 360°]        \\
\hline
HO procedure  & UL RS periodicity   & 160 ms      \\
          & UL-offset           & 1 dB       \\
              & UL-TTT          & 32 ms       \\
\hline
Secure NC  & Field size     & $2^8$       \\
          & Packet size    & 1024 bytes  \\
          & Generation size    & 32      \\
\hline
\hline
\end{tabular}
\end{table}

\subsubsection{Secure NC Related Assumptions} \label{security assumptions}
Different BSH controllers in the simulation scenario are considered as different security domains. In other words, each BSH-cell has its own set of security keys used for secure NC-enabled communications inside the BSH-cell. We consider an RLNC-enabled network in which D2D communication is supported. The finite field size of $q=2^8$, packet size of 1024 bytes, and generation size of 32 are considered as the network parameters unless specified and these assumptions are summarized in Table.~\ref{table1}. The proposed blockchain overlay is enabled by BigchainDB \cite{bigchainDB2}, a distributed and decentralized ledger with blockchain properties. BigchainDB uses Tendermint Byzantine Fault Tolerance (BFT) for achieving consensus within the decentralized network, providing an immutable decentralized data storage. Further, it also provides quick verification of blocks and achieves consensus much faster compared to many other blockchain models. The bigchainDB server is hosted on a local machine and Postman API collaboration platform is used to enable web query based communication with the server. 

\subsection{Comparison of Different Schemes} \label{comparison}

This subsection presents the comparison and analysis of the proposed approach with different schemes in the literature. This analysis done for both the security of the approaches as well as the signaling cost. The blockchain-based security and handover control ensures a secure distributed mobility management for NC-enabled small cell environments. The blockchain-based dynamic mobility management schemes \cite{sharma2018secure} are already known to address hierarchical security issues in edge and fog networks and our proposed approach can also be compared in this aspect and ensures there is no single point of failure and no hierarchical dependencies. The distributed key management scheme proposed allows a new node to join the network anytime after proper authentication performed by the BSH controller and any existing node can also leave the network without effecting other participants. This ensures that the proposed key management scheme has proper forward and backward security. Furthermore, in case of an existing node moving from one cell to another, the proposed approach ensures the keys required for the integrity schemes are shared through the distributed ledger, keeping the security of the network intact. However, the proposed key management protocol is used only for the integrity schemes currently, not for general authentication of participating nodes, making the security analysis to be more focused on the defense against pollution attacks. Additional applications of the proposed approach for authentication and further security analysis of the proposed approaches are to be considered as the possible extensions of this work.

\subsubsection{Security Against Pollution Attacks} \label{security}
The level of security against pollution attacks using MACs will depend upon the field size used for the operations as well as the number of tags/keys used in the integrity scheme. In most of the practical applications of RLNC, a field size of $2^8$ is used. If $q$ is the field size, then a single tag attached to the packet will ensure that the probability of an adversary successfully introducing a polluted packet to the network and pass through the tag verifications is $1/q$. Practically, this security may not be sufficient for all the applications and we generally use multiple tags to increase the security of the integrity scheme. If $l$ number of tags are attached to the packets, then we can achieve a security level of $1/q^l$ with a bandwidth overhead of $l/(m+n)$ over the coded packet where $m$ and $n$ are generation size and packet size respectively. Thus in a field size of $2^8$, a single tag attached to the packet provides a security level of $1/2^8$ against pollution attack. 

Another main factor that affects the security of the integrity scheme and the number of tags is the probability of colluding adversaries. Multiple adversaries may cooperate to successfully bypass the integrity check. Most of the previously existing integrity schemes addressed this challenge by employing specific key distribution schemes so that the participating nodes may not have all the keys used by the source node to create tags. However, this method has multiple drawbacks and scalability issues. In such cases, the number of tags attached to each packet will also depend on the probability of colluding attackers. For example, in a c-cover free set system based key distribution, the source node should have at least $c$ times the number of keys than any other participating node to achieve security against $c$ colluding attackers \cite{canetti1999multicast}. 

In a dense cell environment, the probability of colluding attackers may be very high and it will also increase the number of tags required to provide sufficient security. This probability will depend on the authentication protocols followed in the network and security standards of the participating devices. Furthermore, the overhead due to these increased tags will not provide equivalent security to the number of tags attached, but only equal to the number of tags that can be verified at a particular node. In other words, even if $L$ number of tags are attached to the packets by the source node and transmitted across the network, a receiving node that holds only $l$ keys can verify only that many tags and thus provide a security level of $1/q^l$ only. This results in a mismatch between the bandwidth overhead of the system and the security level. 

Our proposed approach addresses these problems by sharing the tags through not only the communication channel but also the blockchain so that an adversary can not modify the packets and create valid tags even if they have all the secret keys. In our scheme, we consider a strong adversary that can possess all the secret keys that are used for tag creation. However, since the original tags are shared directly by the source node to the blockchain, an adversary can not modify the tags that are stored in the blockchain. Thus even if it creates valid tags for a polluted packet, the next genuine receiver will discard the packet since it will not match the corresponding tags stored in the blockchain. This situation does not differ even if multiple adversaries are colluding to bypass the integrity check. Since our integrity scheme does not require a key distribution protocol that depends on the number of adversaries in the network, we can allow all the users to have the complete key set and verify all the tags that are attached to the packets. Thus in our key management scheme, all the keys in a security domain will be available to all the participating users in that security domain. This will also allow the system to achieve a high-security level solely depending on the number of tags attached to it. Furthermore, the key management scheme is easier to implement since all the keys can be provided to any new incoming node. 

The security level achieved by all the integrity schemes \cite{adat2018blockchain,zhang2011padding,esfahani2017efficient} per node is the same if the same number of tags are verified at that node. However, for other comparable integrity schemes \cite{zhang2011padding,esfahani2017efficient}, the key distribution should happen through a random distribution. In these schemes, the number of keys in total and the number of keys provided to a participating node also depends on the number of colluding adversaries in the network. More specifically, the number of tags required to provide at least $1-\epsilon$ probability of having $c$-secure key distribution in the network is defined by the equation \ref{MacSig equation} where $d$ is a security parameter.
\begin{equation} \label{MacSig equation}
    \displaystyle L=\frac{1}{(1-d)}e(c+1)ln\frac{1}{\epsilon} .
\end{equation}

The bandwidth incurred by $l$ number of tags for \cite{zhang2011padding} is presented in (\ref{MacSig_B}) and for \cite{esfahani2017efficient} is shown in (\ref{HMAC_B}) where $m,n,q$ are the generation size, packet size and field size respectively. 

\begin{equation} \label{MacSig_B}
    \displaystyle B_{MacSig} = \frac{l+1}{m+n} + \frac{32l}{q(m+n)} .
\end{equation}
\begin{equation} \label{HMAC_B}
    \displaystyle B_{HMAC} = \frac{l+1}{m+n} .
\end{equation}

\begin{figure}[!b]
\includegraphics[width= \columnwidth]{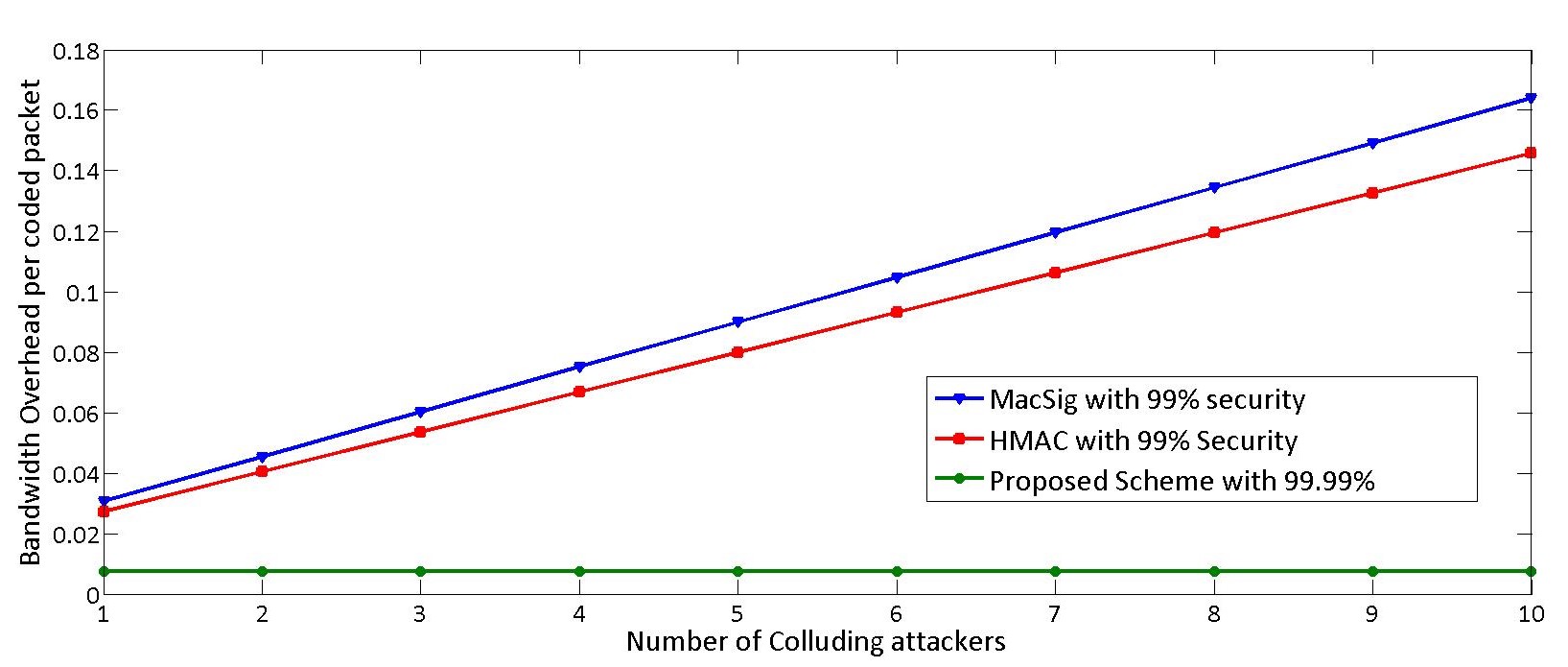}
\centering 
\caption{Bandwidth overhead at comparable situations.}
\label{fig4}
\centering
\end{figure}

Since our integrity scheme allows every node to have the complete set of keys, the number of tags that can be verified at a particular node is maximized. If other integrity schemes require to achieve the same level of security, they need to have a larger number of tags attached to the packet and more keys stored at the source node. Fig.~\ref{fig4} shows a comparison of the bandwidth overhead due to the number of tags to achieve the same level of security in different schemes against different numbers of adversaries present in the network. The bandwidth overhead of the proposed approach is very low and remains the same irrespective of the number of colluding attackers while the other schemes incur a steady increase in the bandwidth overhead with an increase in number of colluding attackers. This is because of the dependency of the other schemes such as \cite{zhang2011padding,esfahani2017efficient} on the number of keys that can be distributed to the participating nodes against colluding adversaries as explained with eqations \ref{MacSig_B} and \ref{HMAC_B}. If there are more adversaries in the network, these schemes require more tags to be attached to the packets increasing the bandwidth overhead. Our proposed approach allows each participant to have all the keys, thus do not require such a gradual increase in the number of tags to defend multiple adversaries. This ensures that the bandwidth overhead remains low irrespective of the number of adversaries in the network with the proposed approach.

\begin{figure}[!t]
\includegraphics[width= \columnwidth]{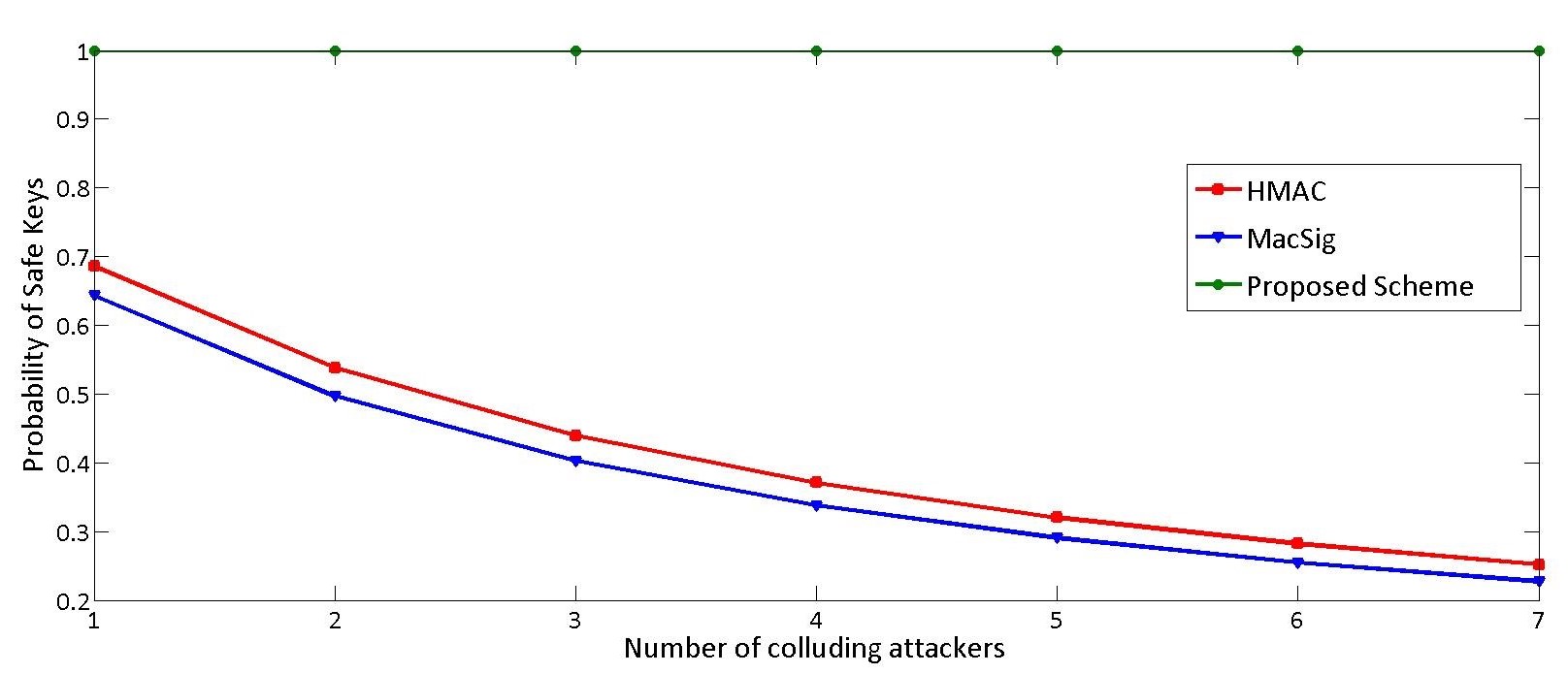}
\centering 
\caption{Probability of safe keys in the network.}
\label{fig5}
\centering
\end{figure}

On the other hand, the probability of having a safe set of keys in the presence of colluding adversaries if the bandwidth overhead due to tags is made to constant is shown in Fig.~\ref{fig5}. In the proposed scheme, this probability of having safe keys is fixed to the probability of randomly guessing the valid keys while in the other comparable schemes this is inversely proportional to the number of colluding attackers. This also means that the probability of safe keys in the network remains unaffected by the number of colluding adversaries and remains close to 1. However, this probability will depend on the number of keys used in the integrtiy scheme. If we use only one key, the probability will be $1-(1/q)$ and if we use $l$ keys, the probability of safe keys will be $1-(1/{q^l})$. We considered 8 tags attached to each packet and the number of colluding attackers varying from 1 to 7 for the analysis shown in Fig.~\ref{fig5}.  

\subsubsection{Signaling Overhead} \label{signalling}
In case of a UE moving from one BSH-cell to another one, the set of keys used in the destination cell needs to be shared with the node during HO. In traditional approaches, like double random key distribution or c-cover free based key distribution, only a random subset of the keys can be shared with the incoming node. This requires the t-BS to select a specific set of keys to be shared with the incoming node and pass it to the s-BS. This process is triggered by every HO request. In other words, for every HO request, two signal exchanges (one from t-BS to s-BS and another one from s-BS to UE) are required in both double random key distribution and c-cover free based key distribution. On the other hand, in our integrity scheme, the signaling overhead for key distribution is different for the initial and subsequent HOs as explained in section \ref{Uplink HO}. 

In our proposed scheme, the keys are shared through the bigchainDB. In this approach, for the initial HO, the key sharing involves three signals, one from t-BS to the bigchainDB server sharing the candidate block including the keys, second one from the server to other bigchainDB nodes including s-BS with the verified block, and a last one where the keys are shared to the corresponding user from s-BS. On the other hand, the subsequent HOs to that particular t-BS will not require the initial two transactions since the keys are already stored in the distributed database. It only requires the s-BS to provide the required keys to the user moving to the t-BS. Thus, the signaling overhead in our approach is 3 transactions for the initial HO and one for the subsequent HOs. Further, the signaling for updating the ledger in each node includes all the verified blocks within that period and if multiple HO requests are handled during that time, only a single communication from the bigchainDB server will serve all those HO requests. It has also to be noted that the initial three transactions are required only for the first HO request to a particular t-BS irrespective of the s-BS since all the BSs are part of the distributed overlay and will receive the verified block during the process. 

\begin{figure}[!t]
\includegraphics[width= \columnwidth]{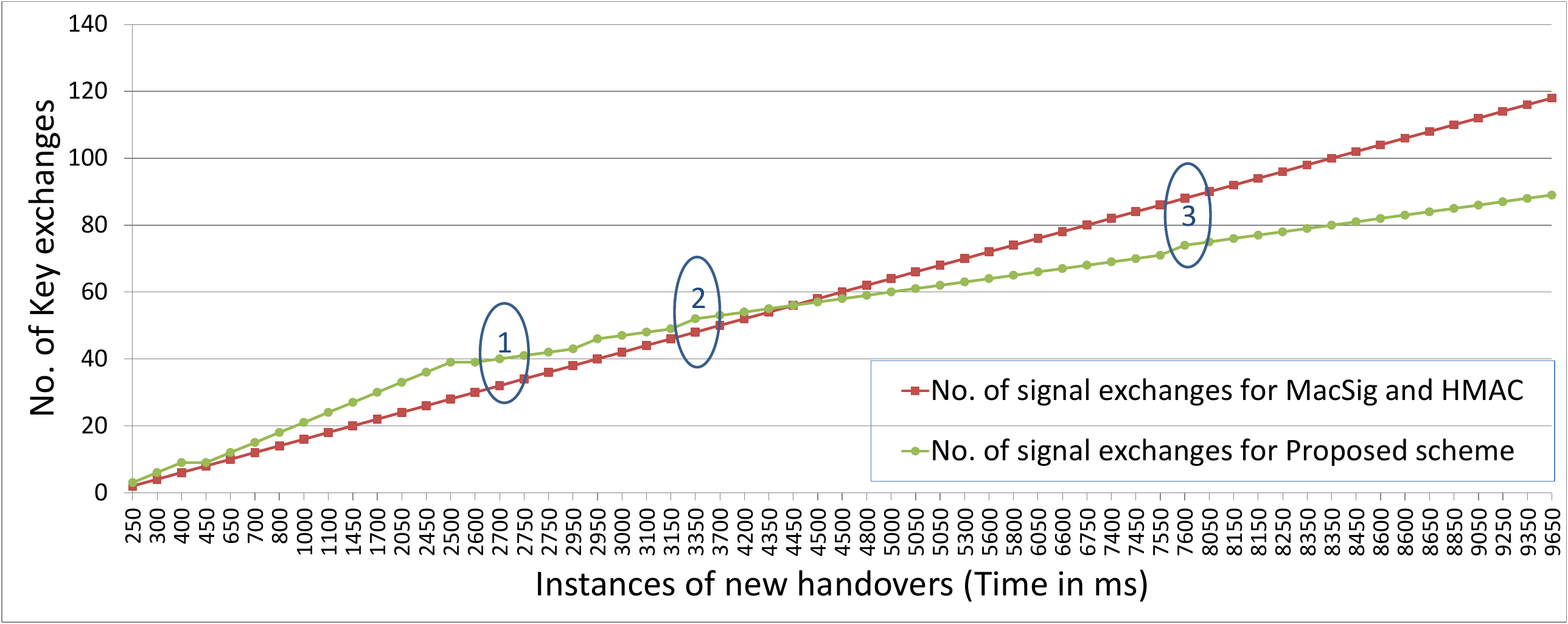}
\centering 
\caption{Number of key exchanges for different schemes.}
\label{fig6}
\centering
\end{figure}

The benefits of our integrity scheme in terms of signaling is analyzed as per the simulation scenario provided in section \ref{simulation assumptions}. The number of signal exchanges of the different key distribution schemes are shown in Fig.~\ref{fig6}. It can be seen that the new blockchain-based approach outperforms the existing key distribution schemes and, as time progresses, the curve levels out. As depicted in the graph, the proposed key management scheme requires a higher number of signal exchanges in the initial phase since each initial HO requires three transactions in the proposed scheme compared to two in the traditional approach. However, at 2700 ms, we have the first beneficiary of the proposed scheme when a new UE is moving to a small cell that has already stored its keys in the blockchain. This happens as a result of the s-BS already having the keys required by the UE and these keys can be immediately passed to the UE in a single signaling transaction. 

The marked point 2 shows when 15 out of the 16 BSH-cells already stored the keys in the blockchain and any UE moving to these 15 BSH-cells can directly get the required keys from their source BS as explained in the HO completion phase of Fig.~\ref{fig2}, requiring only one signal exchange against the two signal exchanges required in the other schemes. It can be seen from the graph that very soon, the total number of signal exchanges of the proposed approach becomes lower than the other existing approaches. The last marked point in Fig \ref{fig6} shows the instance where all the small cells have received at least one new user and thus stores their keys in the blockchain. After this point, any HO in the proposed scheme will require only a single signal exchange between the source BS and the UE for the key exchange and from this point each handover will require only half the number of signal exchanges in the proposed approach compared to the other comparable schemes. This graph is prepared based on our simulations for 10 s but has not considered the delay due to block validation. A detailed study considering the latency and a practical scenario for key sharing is discussed in the next subsection and presented in Fig.~\ref{fig7}.

\subsubsection{Latency} \label{latency}
The blockchain-based key distribution introduces some latency in the key sharing process due to the block verification and consensus process. In the traditional approaches, the key sharing was a direct decision by the t-BS and there was only a procedural delay to choose a random set of keys or a carefully chosen subset of keys for the incoming node. However, in the blockchain-based approach, the keys need to be sent as a candidate block to the server and then verified and added as a verified block in the chain. Due to the security requirements of block verification and reaching consensus, this process introduces a delay in the key exchange.The verification time between two blocks is called the collection period and all the transactions that are received during this time will be added to a single block in the blockchain. In bigchainDB, the minimum time required between the verification of two blocks is one second. This results in a maximum delay of 1 s for a key sharing during HO and introduces the latency in the network. However, at every block verification, all the valid candidate transactions will be verified and added to the chain in a single go. This implies that if there are multiple UEs requesting HOs during the collection period, then all of the required keys will be served to them once the next block is verified. From a signaling perspective, this will reduce the number of signals from the bigchainDB server to other participating nodes (BSH controllers) to a single broadcast of the verified block per second.  In this case, after every block verification, the nodes will update their local ledger with the newly verified block. Thus in a real case scenario, the per-second signaling ($N_{sig}$) for our proposed key management scheme is the sum of the number of BSH controllers that have to store their keys to the blockchain ($n_{bsh}$) and the number of UEs requesting HO ($n_{ue}$) along with one database update signal as shown in (\ref{signallingeq}): 

\begin{equation} \label{signallingeq}
    \displaystyle N_{sig} = n_{bsh} + n_{ue} + 1 .
\end{equation}

Furthermore, the UL-HO scheme helps to predict the possible handovers in advance, which can be used to tackle this issue of latency. We have analyzed the impact of this prediction of future HOs with 80\% accuracy to see how this helps to reduce the signaling (We considered the 80\% as a representative average value of the best case of an 100\% accurate prediction and a lower acceptable rate of 60\%). This analysis shows that with the help of UL RS based prediction of HOs, the proposed key management scheme may have a slightly larger number of key exchanges initially, but a cumulative analysis of the number of key exchanges shows that this prediction helps to achieve a smaller number of key exchanges and thus improve the energy efficiency. This is happening because of the larger number of BSH-cells are required to exchange their keys with another cell initially and this results in larger number of signals being exchanged in the initial instances. However, once a BSH-cell has registered its keys in the blockchain, any further incoming nodes can get the keys with a single signal exchange with its s-BSH making the cumulative number of signal exchanges to go lower than the non predictive scenario very quickly. 

\begin{figure}
\includegraphics[width= \columnwidth]{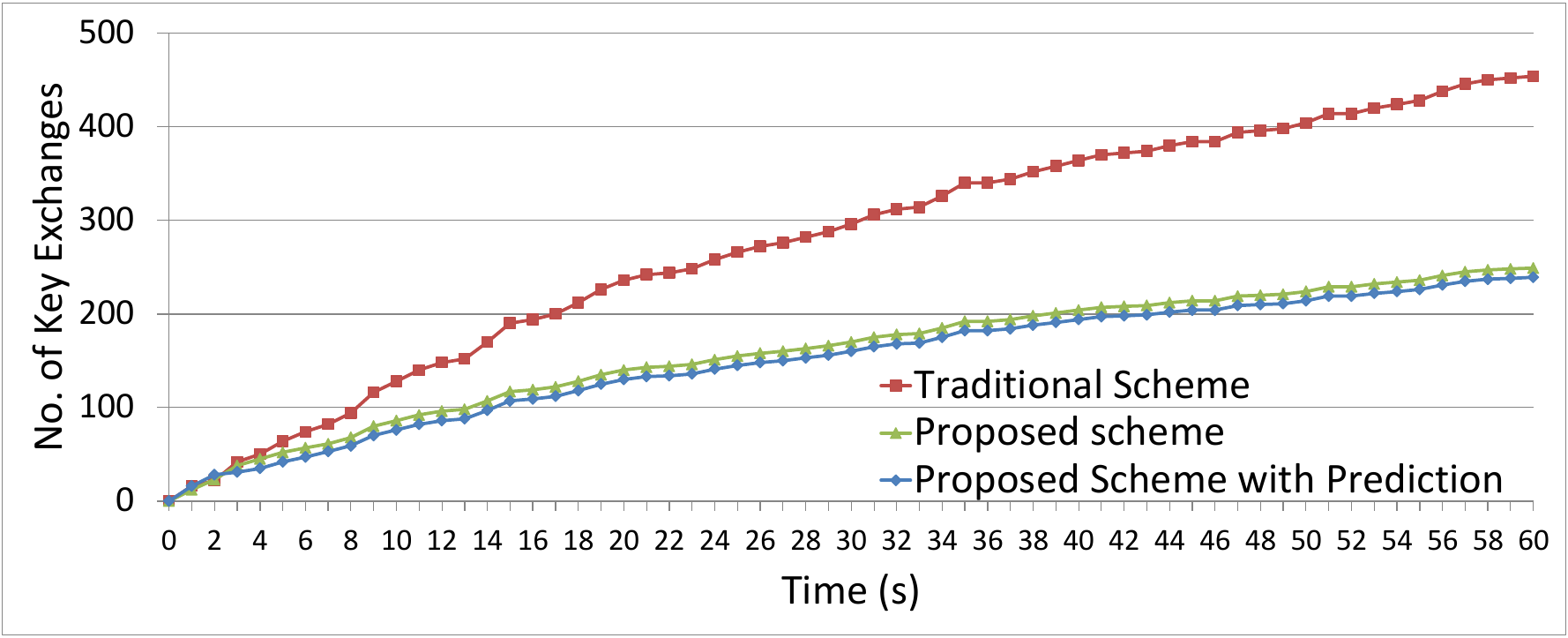}
\centering 
\caption{Cumulative key exchanges for extended time span.}
\label{fig7}
\centering
\end{figure}

\begin{figure}
\includegraphics[width= \columnwidth]{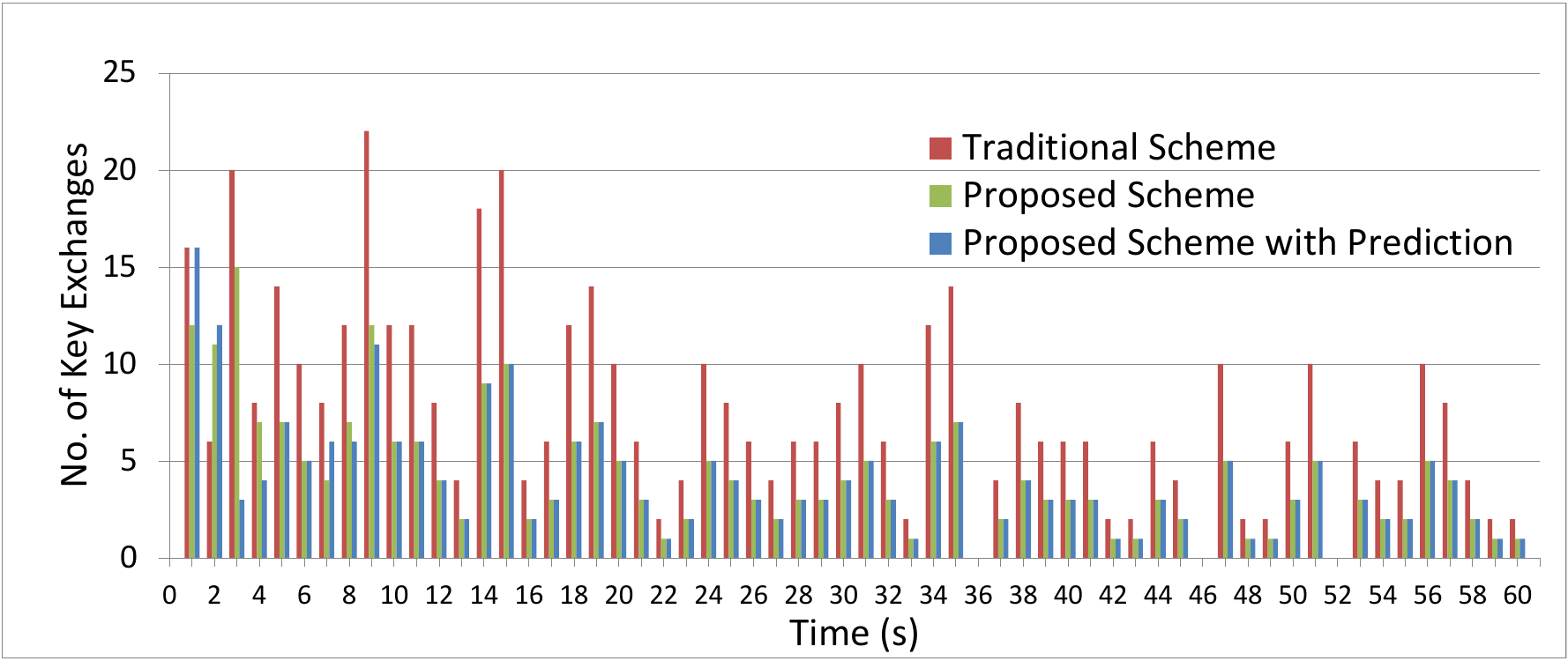}
\centering 
\caption{Number of key exchanges per second.}
\label{fig8}
\centering
\end{figure}

In both cases, the number of signal exchanges remains almost half of the number of signal exchanges required in traditional approaches clearly showing the improvements of the proposed key management scheme. Figure \ref{fig7} shows the cumulative number of signal exchanges for a 60 s time frame considering the block verification delay with and without prediction and Fig.~\ref{fig8} shows the per second signal exchanges for the proposed scheme with and without prediction compared to the traditional key sharing schemes. Although the proposed scheme introduces a latency, in a real scenario, the signaling costs are considerably reduced by the proposed blockchain-based key sharing with the UL-HO scheme. Furthemore, only the first incoming HO will require three signal exchanges and any further HOs will require only a single key exchange in the proposed approach. On the other hand, the traditional schemes will require two signal exchanges in every HO. This also indicates that an extrapolation of the scenario to larger number of small cells and users will result in better results with the proposed approach. 

\section{Conclusions} \label{conclusions}
Since the 5G and beyond systems are expected to deploy dense small-cell networks, NC-enabled small cell environments with D2D communication is beneficial for a large majority of applications ranging from high-speed content distribution to virtual reality applications. Ensuring security while the participating nodes move around multiple small cells and addressing the challenges of mobility requirements with energy efficiency are among the major constraints to be addressed to enable such scenarios. As we focus on the integrity schemes for NC-enabled environments, the keys used for MAC creation and verification needs to be properly distributed in the network. Even if a new node enters the network or an existing node moves from one security domain to another, it should be assigned with the required keys to be part of the secure network coding environment. This paper describes a distributed key management protocol which address the issue of key sharing for mobile nodes.

The proposed integrating approach combining a UL-HO scheme and blockchain-based key management protocol ensures an energy efficient secure NC-enabled small cell environment. We observe that the overheads due to the blockchain-based key management scheme are significantly smaller compared to the traditional key management approaches and the UL-HO scheme has better energy efficiency compared to the legacy HO schemes. The analysis of the integrated approach using the concept of BSH controllers shows that it provides higher security against colluding adversaries and achieves a high level of security with a lower number of MACs and also reduces the signaling overhead during the HO process. The signaling during the HO process becomes linear with respect to the number of HO requests. As the time progresses, it further reduces by half the number of signal exchanges required when compared to other key management schemes. This proposed approach ensures that the integrity scheme for secure NC-enabled next generation networks can work flawlessly in a mobile environment.

  \section*{Acknowledgment}
This project has received funding from the European Union's Horizon 2020 research and innovation programme under the Marie Sklodowska-Curie grant agreement No 722424

\bibliography{references}

\end{document}